# Evolving Categories

## Consistent Framework for Representation of Data and Algorithms

Evgeny Yanenko

A concept of "evolving categories" is suggested to build a simple, scalable, mathematically consistent framework for representing in uniform way both data and algorithms. A state machine for executing algorithms becomes clear, rich and powerful semantics, based on category theory, and still allows easy implementation. Moreover, it gives an original insight into the nature and semantics of algorithms.

## Introduction[1]

Since famous Turing machine we have learned a lot about algorithms, but much more about data structures. I think the main reason is following: data has a meaning, it describes something from the real world, or one says, it has semantics. Algorithms, on the other side, describe mystic transitions of mystic state machine depending in mystic way from such understandable data.

One can see how good we find, sort and reuse data now – we easily handle huge databases and we find a necessary piece of information in Internet. Of course, handling of data is still far from ideal. But there is also a lot of effort to improve it, for example, to order the information on Internet – just to mention semantic web with XML (eXtended Markup Language) and RDF (Resource Definition Framework).

Of course, there is also progress in developing and reusing algorithms. One can use standard libraries, components, etc. And it works, until one needs, for example, to change some trifle in the ready component.

Compare it with changing a sentence in the document.

Often it is easier to write a new component from scratch. I would be not so wrong, if I state, that, at the moment different programmers perform essentially the same work not even twice – thousands and tens thousands times.

So, there is a little, if any, progress in systemizing algorithms. There exist a lot of concepts, used in programming languages, databases, expert systems, etc., but the only thing, which unites them, is Church's lambda-calculus and Turing's machine.

An interesting idea was proposed by Yuri Gurevich with an attempt to represent algorithms as "Evolving Algebras" [1]. To my opinion, author has selected a false start point, describing algorithms separate from data, and used inappropriate semantics carrier. Special algebras are not transparent enough to be used to describe sense of algorithms. But this article has delivered a last puzzle piece into the suggested framework and also helped to give it an appropriate name.

My aim was to understand semantics of algorithms in terms of semantics of underlying mystic state machine and its mystic transitions. Such approach may be useful to find a common basis for different programming approaches.

## Structure of state machine

Turing's state machine is a perfect theoretical construct, but it is absolutely unusable in praxis, with the exception to torture students.

Modern computers use state machines with different instruction sets, but only handful experts can express their thoughts in these terms.

Lambda-calculus delivers a way to define syntax to write down recursive programs. Extensive research on different types of lambda-calculus has permitted to specify a lot of different, so called, high-level programming languages, like C, Java or Pascal, which we use to explain our intentions to computer. To translate such programs into the set of commands, which computer can execute, we use compilers. Programming languages are quite understandable, if it is your own program. But it is really a challenge to understand a program, written by someone else.

As an extreme example, one can have a look to some program, written in ABAP, programming language used by SAP. There were developed thousands of such programs, which work till now and it seems to me, SAP has to maintain an ancient language, because nobody is able to understand what the working programs do.

To solve the problem one can try to create a new, very transparent and extremely high-level programming language, but still it will be syntax for writing algorithms. I think there is a better way. One should start at the roots.

> Let us create a state machine, which has sense.
>
> And understand semantics of algorithms through the clear semantics of state machine.

The state of the machine will be represented with usual tree. I will call it just state tree. Each edge (arrow) of this tree has a label and each node can have an arbitrary number of children. The path to the node, starting with the root, uniquely identifies each node. So, nothing special: everybody knows this kind of addressing. Here are some examples:

http://www.math.ucr.edu/home/baez
C:\Windows\System32\Restore\rstrui.exe
Yanenko@t-online.de

---
[1] Published on 18.12.2004 at http://www.arxiv.org

*Germany, 98765, Irgendeinstadt, Irgendeinestrasse, 2*

Now it is necessary to define transitions from one state to another. And each such transition comprises following steps:

*Build some new tree, according to the next instruction and current state tree. One can also take into account some external events or conditions, e.g. the weather outside.*

*Calculate some node in the state tree, according to … (see above).*

*Replace a sub-tree, starting in just calculated node with the new tree.*

The following picture shows such transition.

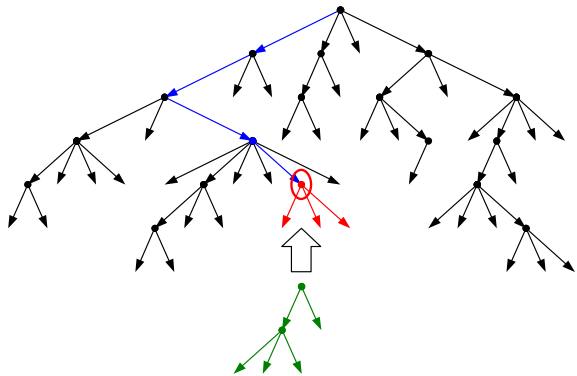

Picture 1. Transition between states.

Such kind of transition appears to be also very familiar. For sure it is easy to find much more examples from the real life:

*Sorry, our site has moved to http://www.newaddress.com*

*We have married and our new address is: New street, 17*

And some in pictures:

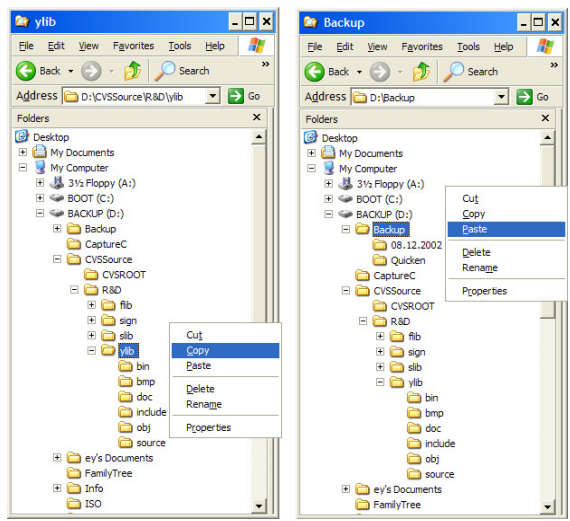

Picture 2. Copy and paste.

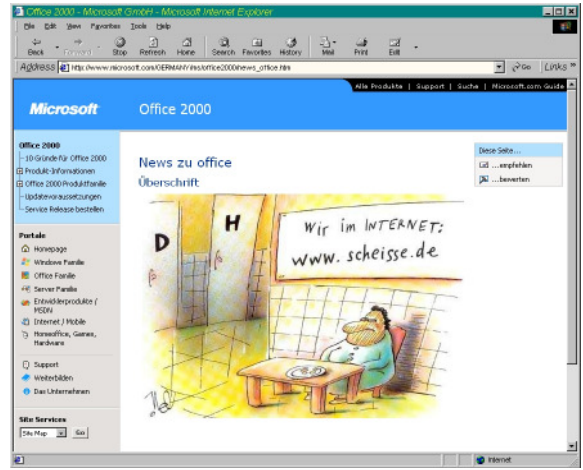

Picture 3. And this happens as the result of unauthorized transition, done by some hacker.

## Possibilities of state machine

Now the draft of the state machine is ready. Of course it is necessary to define an appropriate instruction set, but it will evolve later and quite naturally from the semantic structure, which will be added to the state tree.

There is nothing new or special. Such machines are used elsewhere and they already allow to perform a lot of difficult and useful tasks. They are extensively used in computing. For example, next picture illustrates an algorithm for calculating arithmetic expressions.

Each expression will be represented in computer as a tree, with nodes, representing operations, and leaves, representing arguments.

Each transition looks for the operation, which is possible to perform and replaces sub-tree in operation node with the result of performed operation.

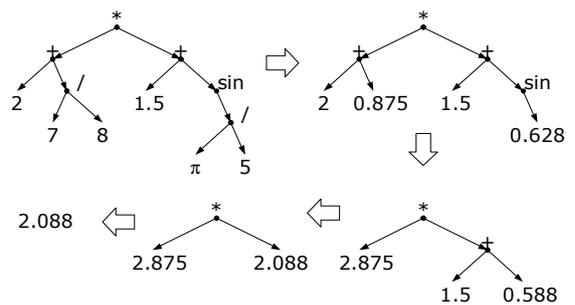

Picture 4. Calculating $(2+7/8)*(1.5+\sin(\pi/5))$.

Moreover, this approach can be extended to implement a powerful programming language, like LISP.

Another example is the relational database, which also has a tree-like structure with the possibility to insert, delete, replace nodes, and to create new trees, containing results of database queries.

A file system in the computer, as well as whole Internet have also a hierarchical tree structure with the possibility to create and replace sub-trees.

What is then wrong with this very well known and so useful state machine?

The answer is simple – *it is too abstract*. It can describe too many different things, which have nothing to do with each other.

It has no internal structure. It has no semantic.

I will start also with very abstract notion – category and slowly add more and more structure to the state machine.

## Applying structure – addressing in the tree as category

Most researchers use category theory as a model for different types of lambda-calculus. A fundamental connection between typed lambda-calculus and Cartesian closed categories is well known in computer science, see e.g. [5], [8], [10]. Category theory was successfully applied in data structures type theory [6], [7], [13] and in investigations on higher order lambda-calculus [9]. A categorical abstract state machine to execute a certain type of lambda-calculus was also introduced [7].

My intention was to use categories to describe the structure of the state machine itself, instead of representing different models of lambda-calculi.

First, let us consider some definitions from category theory with a simple example to get a feeling about the matter. I will try to give short definitions of terms before I use them. Those, who are familiar with the theory, can skip this.

Category theory deals with a set of objects, which I will denote with capital letters, like A and B, and arrows between objects, denoted with small letters, like f: A→B, or arrow f from A to B. Operations, which map arrow f to objects A and B are denoted as A=dom(f) and B=cod(f).

Additionally following axioms are stated:

> *For each pair of arrows g: A→B and f: B→C there exists an operation f∘g = h: A→C. Arrow h is called composition. Associativity is required: (f∘g)∘h = f∘(g∘h).*
>
> *For each object B there exists an identity arrow $1_B$, so that f∘$1_B$ = f and $1_B$∘g = g.*

Consider following example. Let composition operation be "/" or "\", just as one likes. A composition of arrows "etc" and "bin" then will look like "etc/bin". All identity arrows will be called ".", so "bin/." = "bin". It is also possible to build an arrow from the root of the tree to any node: "/etc/bin".

One can also use natural numbers as arrow labels, and dot as a sign for composition. One can assign labels 0, 1, … from left to right to all arrows starting from the certain node. 5.2.3 will then be a blue arrow, shown on the next picture.

Note that there is a special object in this category – root. There exists exactly one arrow from the root to any node.

Such object is called initial object of the category and is denoted with **0**. A unique arrow to object B is denoted with $0_B$: **0**→B.

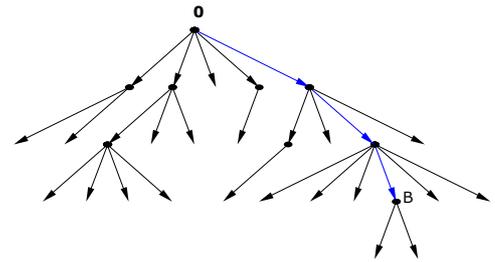

Picture 5. Structure of category **Pos**.

One can imagine, that we have changed the direction of all arrows to opposite. In this case there will be exactly one arrow from each node to the root. Such object is called final object of the category and is denoted with **1**. A unique arrow from object B is denoted with $1_B$: B→**1**.

After changing arrow direction, this category has transformed to another category, which is called dual. It is easy to check, that for arbitrary category there exists a dual category.

As one can see, the state tree can be considered as a category, which I will call **Pos** with initial object and clear semantics. Initial object and composition of arrows allow unique addressing of each node in the tree. Category **Pos** represents our understanding of the physical address – certain object can occupy exactly certain place.

## Extending to Finset – category of finite sets

As far as we know, our world is infinite. Why one should restrict the consideration to the category of finite sets?

It makes things much simpler. There is a huge difference between arbitrary big and infinite. We will never deal with infinite amount of information. Although it will grow bigger and bigger, it will always remain finite.

In general, to simplify implementation of the state machine, I will try to avoid pure mathematical abstractions, like infinity or category of all categories **Cat**, although all categories, described here are objects in **Cat**.

Set is one of the most examined abstract items in mathematics and has perfect semantics. So, category **Finset** is the best selection to apply additional structure to the state tree. It is really additional, because it will reinterpret the meaning of **Pos** arrows, preserving **Pos** properties as a sub-category in **Finset**. Arrows from **Pos** will be used, as before, for node addressing.

**Finset** deals with sets as objects and mappings as arrows. It complies with additional restrictions, which make the structure of **Finset** much more complex as the structure of **Pos**. One can say it is a set theory in category form. Category, for which these additional restrictions hold, is called topos. Categorical definition of topos is not obvious, so, if possible, I will use well-known definitions of set theory instead.

Following picture shows first step of adding **Finset** semantics to the state tree. Of course, it is not possible to represent all objects and arrows of **Finset** in a tree, because this category contains *all finite sets and all their mappings*. From this point of view **Finset** as category is static. It has no internal dynamics, which is associated with the notion of algorithm. The picture below shows only a diagram of some objects and some arrows from **Finset**. This diagram, as distinct from the whole category, can be considered as dynamic, evolving structure.

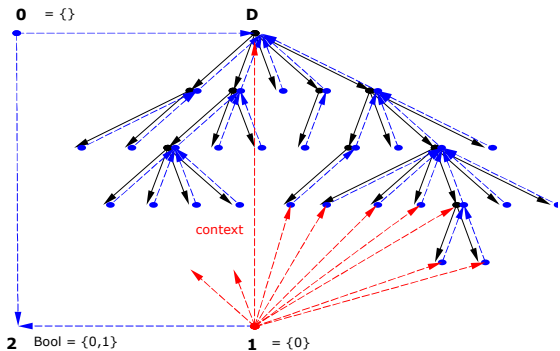

Picture 6. Correspondence of **Finset** objects to the state tree.

All nodes and arrows from the state tree are represented in this diagram. As dashed shown some additional arrows, which are not present in **Pos**. Each node corresponds to some set. Each set can contain other sets as elements. Injection arrows are shown in blue.

Root of the diagram is denoted with **D** and it is a set, containing all top-level sets from the diagram. Of course, it itself is an element of some object in **Finset**.

Leaves of the tree are sets with only one element. Strictly speaking, I have simplified the categorical diagram, using nodes to represent both an element E of the set and the set containing this element {E}.

Non-leave nodes contain as elements objects from **Finset**. To be consistent, elements of leave nodes should also be **Finset** objects. Good candidates are objects from **Finset** subcategory **Finord**. These are **0**={}, **1**={**0**}, **2**={**0,1**}, etc. They are commonly used in set theory to represent natural numbers, because they demonstrate the same properties as natural numbers.

Moreover, **Finord** is a skeleton of **Finset**, i.e. each object of **Finset** with N elements is isomorphic to **Finord** object **N**. Each element of the set can be also considered as a set with one element, therefore all elements are isomorphic to **1**. Corresponding arrows are shown in red. From the practical point of view, a mapping x: **1**→{E}, is an array x with one element x[0]=E. So, exactly these arrows represent actual data. An arrow **1**→**D**, which contains the data for the whole diagram, will be denoted as diagram context arrow.

Empty set, **0**={}, is the initial object of category, but we should also have a possibility to use arrows, starting at virtual root **D**, for addressing. It will be possible, if we interpret **Pos** arrows as mappings from the set to each of its elements. As element is isomorphic to **1**, there exist exactly one such arrow to each element. And this arrow is closely related to injection arrow, which represents mapping of element to its set. As category **Pos** corresponds to physical addressing of the data, it is a good reason to tune physical addressing with logical structure of the data.

There is one more special object in **Finset** – object **2**, set, consisting from {**0,1**} or, in other words, Boolean {true, false}. Such object, which is called sub-object classifier, is a necessary element of topos.

One can find in **Finset** many other interesting objects to be used as elements of nodes, but this is not essential at the moment. Note only, that strings, for example, can be represented as sets of natural numbers.

### Each set as a separate diagram

I have tried to reuse or, better to say, reinterpret arrows from **Pos** in **Finset**. It is not a coincidence – to my point of view, it gives a lot of fundamental advantages.

Each sub-tree of **Pos** has absolutely the same structure as the whole tree. *And this is valid for **Finset** diagram also.* Therefore each set can be treated separately as a **Finset** diagram of its own subsets with the root of the tree corresponding to set itself, see next picture.

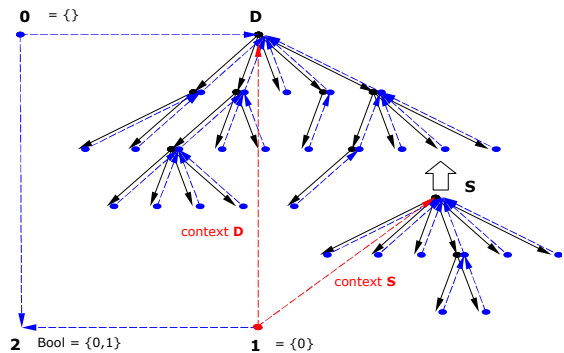

Picture 7. Each set as own diagram with own addressing.

This feature ensures modularity and scalability. Each node can be treated as an independent state machine with own addressing and structure, which can be connected as a separate module to the higher-level machine. So, the whole tree can contain a lot of independent modules.

It is also not necessary to define address arrows in the module starting from initial node – we can speak about addresses in context of diagram **D**, or in context of sub-diagram **S**.

### Operations of category algebra as natural instruction set

Transition of the state machine, which was roughly described at the beginning of this article, can be expressed in the following notation:

<node>=<tree>

where <tree> is a term, responsible for the calculation of the new tree and <node> – a term for locating a node. To specify transitions, it will be necessary to define operations, which can be used in <tree> and <node> terms.

A quite natural choice for building terms is using operations of category algebra. In any category composition of arrows, as well as operations dom(f) and cod(f) can be used.

Following picture shows that for each node B there exists exactly one address arrow and exactly one data arrow. Correspondingly there are two ways to locate a node:

*Node can obviously be located as cod(h∘g∘f). To make this more readable, the node address will be notated as f.g.h, i.e. dot is used instead of a sign for composition and order of arrow labels is swapped.*

*Node can also be found by matching a data arrow. This will be important in the case of term rewriting.*

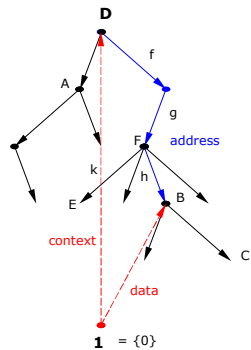

Picture 8. Address and data arrows.

To calculate a data arrow from the address, one will need a composition of context data arrow, corresponding to the root of the tree, with the address arrow:

   data = h∘g∘f∘context

It will also be notated as [f.g.h] – contents of the node, pointed by the address f.g.h.

One can consider address category **Pos** as a category of partial order with relation B ⊆ C = true, if there is an arrow between B and C. Such category with initial object will additionally have a product for each two objects. For example F = E ∩ C – product of E and C is the maximal object, from which E and C are accessible, see the picture above.

To complete the first prototype of the state machine, one needs means to calculate <tree> term.

Topos category structure delivers additional algebra operations, which can be used for these purposes, for example products and coproducts, or, more general, limits and colimits. Following pictures illustrate how these operations on objects and arrows create new trees in **Finset**.

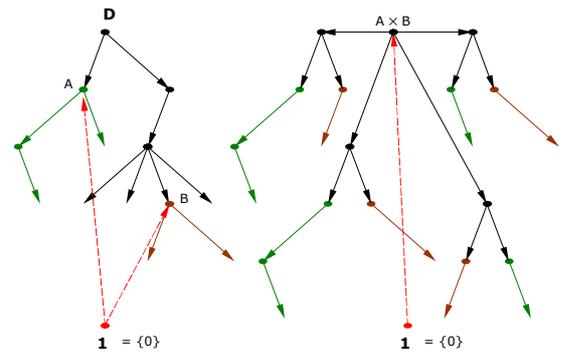

Picture 9. Product A×B.

Product of objects A×B contains all possible pairs <element from A, element from B>. It is just well known Cartesian set product from set theory.

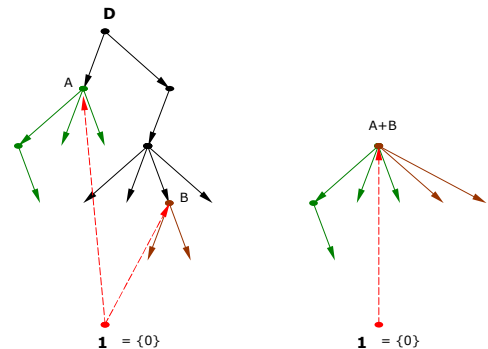

Picture 10. Coproduct A+B.

Coproduct A+B contains all elements from A and all elements from B. It is not a union of sets A and B. A+B contains also duplicates of equal elements in A and B.

As we assumed before, leaves of the tree contain **Finord** objects. For them product and coproduct are just corresponding arithmetic operations.

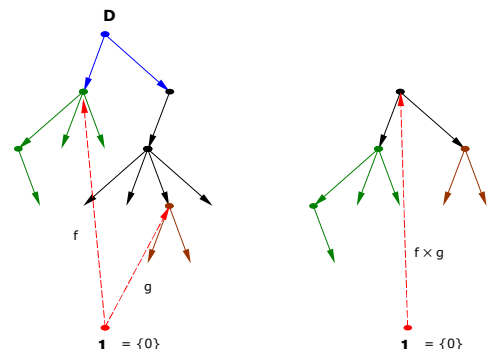

Picture 11. Product of arrows f×g.

Product of arrows f×g is a pair <element f, element g>.

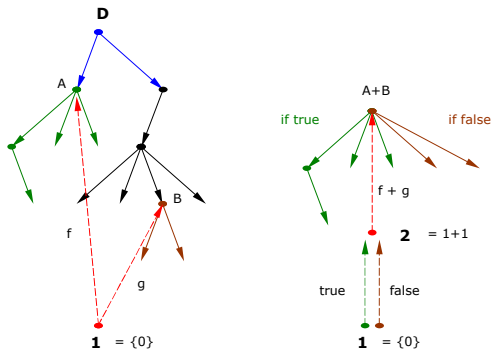
Picture 12. Coproduct of arrows f+g.

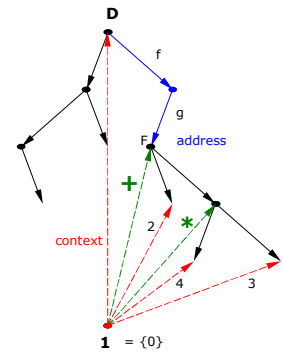
Picture 13. Term representation.

Coproduct of arrows f+g is either f or g, depending on the composition with true or false. It is exactly an "if" construction, which is an obligatory part of each programming language.

### Sub-categories and internal categories

In the similar way, a whole bunch of operations evolves, if one considers sub-categories, like **Pos**, in **Finset**, as well as internal structure of different **Finset** objects as its own category.

Consider, for example, set **N**. It is a category of partial order with relation $A \leq B$ as an arrow, 0 as initial object and N-1 as final object. Such category has both product $A \cap B = min(A,B)$ and coproduct $A \cup B = max(A,B)$. One will also need a supplement $\mathbf{N}\backslash A = N-A$.

Another important set is **2** (Boolean). It can be considered as a lattice, where product and coproduct correspond to usual Boolean operations AND and OR, Cartesian closure $B^A = A \Rightarrow B$ and supplement $\mathbf{2}\backslash A = !A$.

All these operations form a basis of the instruction set for the state machine.

### Terms as tree items

Terms are legal objects, which a set can contain. One can construct a special representation for them as for distinct objects in **Finset**. But, to my point of view, there exists a very elegant solution – to exploit the fact, that each term comprises *a set of operands*, which is already an object in **Finset**. To specify an operation it is enough to assign an operation identifier as a label (like for address arrows) to the data arrow.

Consider an example, shown on the next picture, where data arrow is labeled as a sum. Members of the set, which is represented with this arrow, are summands. Operations, shown on the picture in green, can easily be nested to form a tree analogous to the tree, shown on the picture 4.

Term has to be evaluated at the moment, when a content of the node F is accessed. In the above example it will be a data arrow, but one can also use node terms for references to other objects. To some extent, such terms are similar to spreadsheet formulae, like "=A1*B10+B11", but using, naturally, **Pos** addressing. In the above example the result will be a data arrow, but one can also use node terms for references to other objects.

### Functions

Why do we need a notion of function? In **Finset**, function is an arrow, static mapping of argument from function domain into the function space. One can represent it just as a set of pairs argument-value. But this is too static. It does not correspond to our intuitive understanding of function.

The matter of fact, that one operates only on a diagram from **Finset**, which cannot contain all argument and function values. There should exist some procedure, which will allow calculating function value from argument. In our case, this procedure should create a new result tree from the tree, representing arguments.

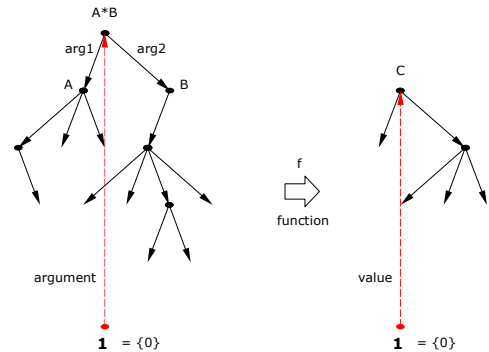
Picture 14. Function f: A×B→C.

It is not clear if Cartesian closure in topos can be "misused" for categorical description of algorithmic function. From the first glance it looks quite promising: 'f': $\mathbf{1} \to B^A$ contains a set of instructions and ev: $A \times B^A \to B$ performs corresponding steps, using g: $\mathbf{1} \to A$ as an argument. But more detailed analysis shows, that in this case there should exist a bijection between arrows f and 'f'. Therefore, all sets of instructions 'f': $\mathbf{1} \to B^A$, representing all possible algorithms for calculating f: A→B, should be isomorphic. In other words, there should exist exactly one algorithm to calculate f. To my point of view, this is not true, but I am neither able to prove, nor to disprove this fundamental supposition. Although this is extremely interesting and theoretically important point, the discussion on this matter is far beyond the scope of this article.

A function on the picture above should perform several transitions of the state machine. Taking into account, that set of instructions comprises pairs of terms, the function body can be also represented in the state tree as *a set of pairs <node term, tree term>*.

**Node location by address**

As it was mentioned above, there exist two methods of locating a node – address matching and data matching.

Consider first address matching: address term is evaluated and the unique target of address arrow is replaced with the tree, calculated by tree term.

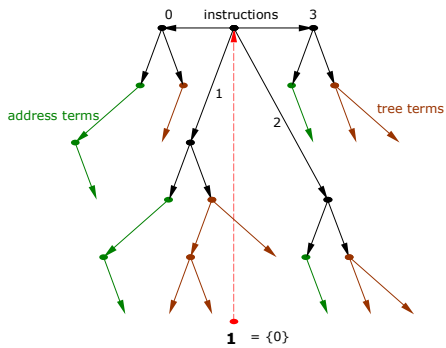

Picture 15. Set of instructions as a tree.

This kind of instruction set is executed sequentially, starting with the first pair. What I miss here is a cycle, like in PHP

*foreach (<set> as <variable>) <instructions>*

Or, at least, jump to instruction number. The latter can be easily implemented with additional variable for instruction pointer. Consistent solution for this item is still open.

**Node location by data**

Data matching leads to essentially different kind of execution, known from LISP or Q, which is called term rewriting. Consider following example, which defines a formula set to calculate greatest common divisor of two natural numbers a and b, a>b.

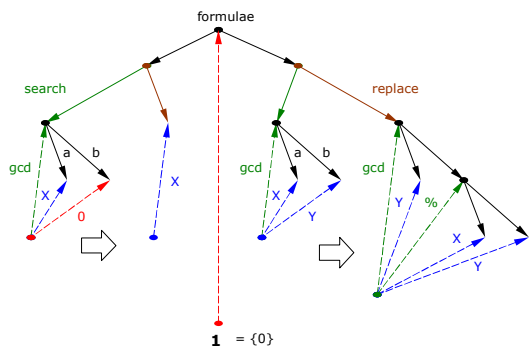

Picture 16. Set of two formulae for gcd(a,b).

Operation % calculates a remainder from division. This diagram represents two recursive equations:

$\forall X \ gcd(X,0)=X$

$\forall X, \forall Y \ gcd(X,Y)= gcd(Y, X \% Y)$

Variables bound with quantor in expressions above are shown in blue. These variables can be matched with arbitrary data.

Execution of some formulae set on some context arrow comprises three repeating steps.

*Calculate all sub-terms with known arguments.*

*Search for all data arrows, which match left part of first formula. If no matches were found – move to the next formula. If it is a last formula – finish execution.*

*Replace found data arrows with data arrows, from the right part of equation, substituting actual found values for all bound variables.*

One can also construct equations, containing terms of the second order. Next formula to calculate a derivative of product of two functions contains functions as bound variables:

$\forall x, \forall f, \forall g$
$d(f(x)*g(x),x)=f(x)*d(g(x),x)+g(x)*d(f(x),x)$

One can see that both types of execution allow building complex algorithms. Address based node location is more suitable for processes, where a certain sequence of actions has to be performed. On the other hand, data driven node search is applicable for rule-based calculations, where execution order is unimportant. Each function should be calculated using only one execution type, but its sub-terms can be arbitrary mixed.

**Function template**

An important question – how to bind function arrow with certain instructions or formulae set. Function semantics is usually closely related to its name, which we use as a data arrow label for the function, and to its arguments, including their names. But often functions with the same semantics perform different actions depending on the current context. So, in any case it makes sense to separate function declaration from function implementation.

The simplest way is just to create a pair declaration - implementation. An example of such pair is shown on the picture below. If necessary, such pair can be repeated in different contexts, using different instruction sets. Note also, that resolution of address terms in the function body will be done in the context of the term, calling the function.

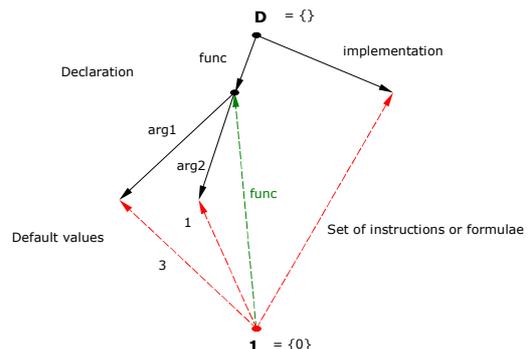

Picture 17. Function.

Simplest solution is not necessarily the best, so the question of binding the function term to its code is a subject for improvements.

To assign a function value to the variable x, one can use following sequence:

```
// Copy function template
f=[func]
// Assign arguments
f.arg1=4
f.arg2=5
// Calculate function term and assign result
x=f
```

Executing a set of instructions should replace the function template with the result of calculation.

It is important to note, that function templates allow implementing of recursive algorithms also in the case of sequential program execution. Function template can be copied by the code of the function itself.

## Types as classifiers

Modern programming languages successfully use types of variables. It is quite logical to use types to classify objects, but in many cases the type is assigned to a label, which makes not much sense. Moreover, it creates problems for reusing semantically correct code with objects of another type. Really, in the nature there is no such thing as type. *Each object is unique.*

Types were invented for object classification. If an object has a certain set of properties, than it can be classified as an object of a certain type. Commonly known example is the classification of plants, animals and insects.

Let M be a set and T – subset of M, consisting of all M elements, for which condition f(x) is true:

$$T = \{x: x \in M \text{ and } f(x) = \text{true}\}$$

In the category **Finset** it corresponds to the limit of the diagram

$$f: M \to \mathbf{2} \leftarrow \mathbf{1} : \text{true}$$

The diagram is shown on the picture below. Such limit is a special case of the limit, which is called pullback. In **Finset** pullback exists for any two arrows g: A→B and h: C→B with the same target.

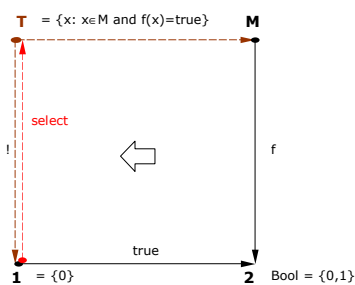

Picture 18. Type as pullback.

As there exists a unique arrow true: **1**→**2**, one can consider a pullback for the pair of functions true: **1**→**2** and f: M→**1** as a tree term, which maps f: M→**1** to the tree data arrow select: **1**→T.

The arrow name "select" comes, naturally from database SQL queries, like:

SELECT * FROM persons WHERE name="John"

Thus, categorical notion of pullback delivers one more tree term for the state machine, which is responsible for selecting objects with certain properties and which allows to perform database-like functions.

## Types as templates

There is one more case, when one speaks about the objects of the certain type – when the objects are produced with the same properties. In this case one can consider a type as a kind of working drawing, like those, which are used by Volkswagen to manufacture Golf.

The next picture shows such working drawing for *Date* type. Note, that along with data fields, the template contains also a member function *weekday()*, which is a mapping from *Date* to a set of {*Monday, Tuesday, …*}.

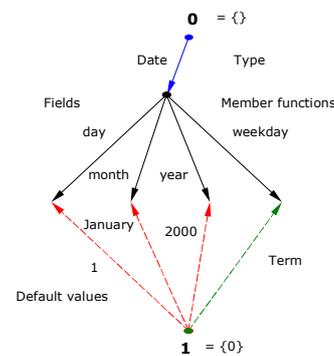

Picture 19. Type as template.

To calculate a weekday for a certain date:

```
// Copy type template
x=[Date]
// Assign fields
x.day=5
x.month=February
x.year=2004
// Calculate day of the week
x=x.weekday
```

Type templates allow working with types just as with usual data. For example, to create a sub-type, it is enough to copy template and to specify some additional fields and functions.

## Appliances

To continue with working drawings, let us introduce appliances. When functions are just mappings from one set to another, the appliances will change the structure of some set in the state machine. Of course, in **Finset** it corresponds to exchanging of one set to another, but such semantics leads to the loss of individuality of the given set. It is easier to understand appliance as changing, evolving set.

Appliances can perform miscellaneous tasks. They are similar to objects in traditional programming languages. I have used word appliance to avoid ambiguity with categorical notion of objects. Again, as with types it is easy to build a working drawing, or prototype of the appliance. Consider an example of heap appliance, shown on the picture below. Heap is often used for sorting data items or as priority queue.

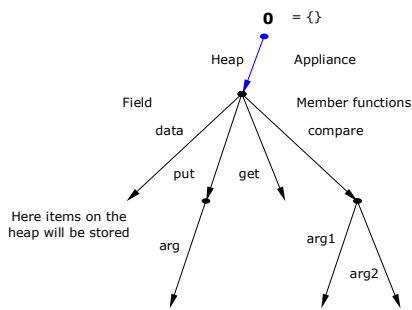

Picture 20. Heap appliance.

This appliance has one set, called *data*, which will be modified with functions *put(arg)* – put some item on the heap and *get()* – remove the top item from the heap. Function *compare(arg1,arg2)* is used to compare items.

To initialize an instance of the heap, it will be necessary to copy heap template and, if necessary, to customize *compare(arg1,arg2)* function.

**Devices**

Devices are necessary part of the state machine, which allow it to communicate with the environment. Input device can be represented with the data arrow, which corresponds to a certain value from the environment – mouse coordinates, key pressed, current time, temperature outside or digitized picture from the camera.

Output device should convert the data arrow, accordingly, into something real – picture on the monitor, sound from the speakers, voltage on the connector, etc.

At the present time the term artificial intelligence (AI) become unpopular, but I'd like to say a few words on the theoretical role of I/O devices for AI. Please do not consider this as an argument pro or contra algorithmic nature of intelligence.

Apologists of the theory of strong AI, which state that AI is just enough complicated algorithm, as well as those, who state non-algorithmic nature of intelligence, see e.g. [4], both underestimate the role of I/O devices. Devices play the role of sense and communication organs of the state machine, which allow evolving an *individual* (for the machine) model of reality, based on the *unique* experience of communication with the environment. They provide a real feedback, even investigation means, which is a necessary attribute of intelligence. Variety and unpredictability of the outside world cannot be coded just as input data on the band of the Turing machine.

**To do**

Several open items were already mentioned. At first, it is lack of categorical description for cycle or jump operation in sequential execution of instructions. And the second – missing consistent polymorphic approach to bind function terms with their instruction or formulae sets.

Still, implementation of the state machine seems to be quite straightforward. One will need three types of labeled arrows – address, data and bound variable arrows together with implementation of basic categorical terms on these arrows. It will also be necessary to define syntax of the language, which is suitable to describe a state of the state machine, as well as a compiler and de-compiler for this language (note: this is not a language to describe algorithms). But most work consuming is to specify state machine representations of programs in different programming languages and to build corresponding compilers.

Till now, I have considered calculation of terms with all arguments defined. An item, which can be investigated as a next one, is matching of terms, containing free variables. In other words, searching for the set of free variable values, which fulfills given equation. This will allow representing, for example, Prolog-like predicate calculus.

Another interesting aspect is extending category **Pos** to the lattice, which reflects the fact, that an object in **Finset** can be a member of different sets. There is also no contradiction between existence of unique physical address and different paths, describing it. One will get five different descriptions how to drive to a certain place from five different people.

**Conclusion**

This work does not pretend to define an exact specification to build "evolving categories" state machine. The intention was to demonstrate a possibility to unite different computation approaches on a single basis.

Suggested concept of "evolving categories", or, more strictly, of "evolving diagrams" allows understanding semantics of algorithms through the semantics of categorically structured state machine. Such structure induces a natural instruction set for the machine, which comprises algebra operations of corresponding categories.

State machine, structured as **Finset** topos, together with categories of partial order, allows to describe consistently in uniform way diverse concepts, used in computing now, starting with data structures through arithmetic and logical operations, data types, functions up to databases and expert systems.

A great variety of different categories can be used to describe structure of certain sets. As an example one can consider formal theories, described as lattices.

But mostly intriguing in the suggested approach is a possibility to replace **Finset** with some *other topos, comprising other, non-Boolean logic*. It is too early to speculate about the perspectives to build a bridge from our binary computation world to non-Boolean computing, but this is a huge subject for further investigations.